\documentclass[multphys,vecphys]{svmult}

\usepackage{makeidx}     
\usepackage{graphicx}    
\usepackage{multicol}    

\makeindex             

\begin{document}

\title*{Constraints on SN~Ia Progenitors and ICM enrichment
from field and cluster SN rates}
\author{D. Maoz \& A. Gal-Yam}
\authorrunning{D. Maoz \& A. Gal-Yam}
\institute{School of Physics and Astronomy,
Tel Aviv University, Israel
\texttt{dani@wise.tau.ac.il, avishay@wise.tau.ac.il}}
%
%

\def\gtorder{\mathrel{\raise.3ex\hbox{$>$}\mkern-14mu
             \lower0.6ex\hbox{$\sim$}}}
\def\ltorder{\mathrel{\raise.3ex\hbox{$<$}\mkern-14mu
             \lower0.6ex\hbox{$\sim$}}}

\maketitle

\abstract

The iron mass in galaxy clusters is about 6 times larger than 
could have been produced by core-collapse SNe, assuming the 
stars in cluster galaxies formed with a standard IMF. Type-Ia 
SNe have been proposed as the alternative dominant iron source.
We use our HST measurements of the cluster SN-Ia rate at high 
redshift to study the cluster iron enrichment scenario. 
The measurements can constrain the star-formation epoch and 
the SN-Ia progenitor models via the mean delay time between the
formation of a stellar population and the explosion of some of
its members as SNe-Ia. The low observed rate of cluster 
SNe-Ia at $z\sim 1$ pushes back the star-formation epoch in 
clusters to $z>2$, and implies a short delay time. 
We also show a related analysis for high-$z$ 
field SNe which implies, under some conditions, a long SN-Ia delay time.   
Thus, cluster enrichment by core-collapse SNe from a top-heavy IMF 
may remain the only viable option.

\section{The SN~Ia Rate in $z \le 1$ Galaxy Clusters and 
the Source of Cluster Iron}
\label{sec:2}

The iron mass in galaxy clusters is about 6 times larger than 
could have been produced by core-collapse supernovae (SNe), 
assuming the stars in the cluster formed with a standard initial 
mass function (IMF; e.g., Renzini 1997). SNe~Ia have been proposed as the
alternative dominant iron source.  Different SN~Ia progenitor models
predict different ``delay functions'', between the formation of a 
stellar population and the explosion of some of its members as SNe~Ia.
We use updated measurements of the total iron mass-to-light ratio in
rich clusters to normalize the predicted SN~Ia rate in clusters vs. redshift,
using the delay function parameterization of Madau, Della Valle, \& Panagia (1998). 
We then use our previous measurements of the cluster SN~Ia rate at high 
redshift (Gal-Yam, Maoz, \& Sharon 2002)
to constrain SN~Ia progenitor models and the
star-formation epoch in clusters.
The low observed rate of cluster SNe~Ia at $z\sim0 - 1$ (Fig. 1) means that, 
if SNe~Ia produced the observed amount of iron, they must have exploded 
at even higher $z$. This puts a $>95\%$ upper limit on the mean SN~Ia 
delay time of $\tau<2$~Gyr ($<5$~Gyr) if the stars in clusters  
formed at $z_f<2$ ($z_f<3$), assuming $H_{o}=70$ km s$^{-1}$ Mpc$^{-1}$
(see Maoz \& Gal-yam 2003 for full details). 
In the next section, we show that, for some current versions of cosmic
(field) star formation history (SFH), 
observations of field SNe~Ia place a {\it lower} bound on the delay time,
$\tau>3$~Gyr. If these SFHs are confirmed, the entire range of $\tau$ will
be ruled out. Cluster enrichment by core-collapse SNe
from a top-heavy IMF will then remain the only viable option.

\begin{figure}
\centering
\includegraphics[height=10cm]{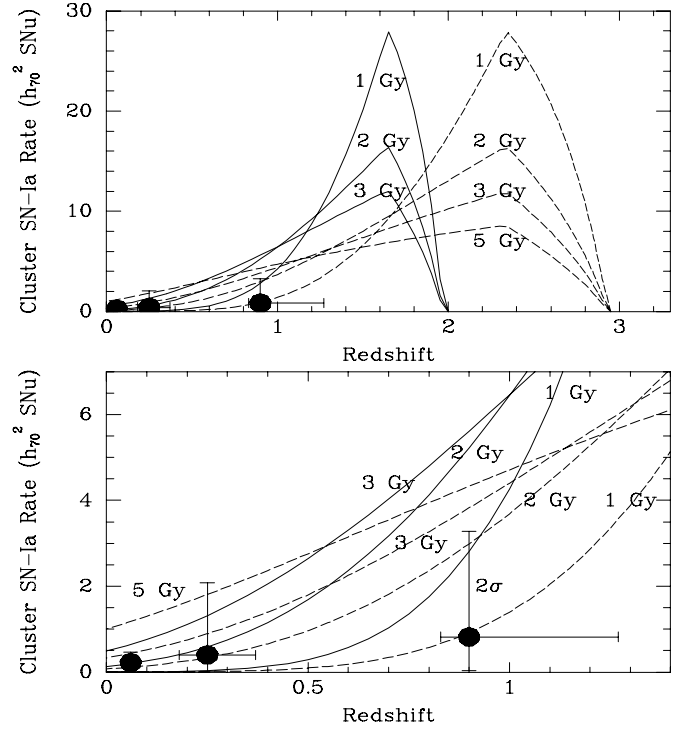}
\caption{Predicted SN~Ia rates vs redshift, if most of the iron 
mass in  clusters is produced by type-Ia SNe following a brief burst 
of star formation at redshift  $z_f=2$ (solid curves) and $z_f=3$ (dashed curves). 
The bottom panel is a zoom on the low-$z$ region of the top panel.
The different curves are for SN~Ia transfer functions with mean delay times,
$\tau$, as marked. Cluster SN~Ia rate measurements are
by Reiss (2000) and Gal-Yam et al. (2002). The latter are shown
with 95\%-confidence vertical error bars. The horizontal error bars
give the visibility-time-weighted redshift ranges of the cluster samples.
The $z_f=2$ models (solid curves) with $\tau\ge 2$~Gyr are clearly 
ruled out by the $z\sim 1$ SN-rate measurement, even after accounting
for a 30\% uncertainty in the nomalization of the models.
The $z_f=3$ model (dashed curves) with $\tau=5$~Gyr predicts unacceptably
high rates at low $z$.} 
\end{figure}

\section{The Redshift Distribution of Field SNe~Ia: 
Constraints on Progenitors and Cosmic Star Formation History}
\label{sec:3}

In this section, 
we use the redshift distribution of SNe~Ia discovered by the 
Supernova Cosmology Project (Pain et al. 2002)
to constrain the star formation history (SFH)
of the Universe and SN~Ia progenitor models. 
Fig. 2 illustrates how, for a given choice of parameters
describing the SFH and SN~Ia delay time, we predict the
observed cumulative (i.e., unbinned) redshift distribution in a given
survey, and compare it to the data. Given some of the 
recent determinations of the SFH, the observed
SN~Ia redshift distribution indicates a
long ($\gtorder 1 h^{-1}$ Gyr) mean delay time between the formation of a
stellar population and the explosion of some of its members as SNe~Ia (Fig. 3). 
For example, if the Madau et al. (1998) SFH is assumed, the delay time $\tau$ 
is constrained to be $\tau \ge 1.7 (\tau \ge 0.7) h^{-1}$ Gyr at the $95\%(99\%)$
confidence level (CL). SFHs that rise at high redshift, 
similar to those advocated by Lanzetta 
et al. (2002), are inconsistent with the data at the $95\%$ CL 
unless $\tau > 2.5 h^{-1}$ Gyr.
Long time delays disfavor progenitor
models such as edge-lit detonation
of a white dwarf accreting from a giant donor, and the carbon core ignition of 
a white dwarf passing the Chandrasekhar mass due to accretion from a subgiant
(e.g., Yungelson \& Livio 2000).
The SN~Ia delay may be shorter, thereby relaxing some of these
constraints, if the field star formation rate falls, between $z=1$ and the
present, less sharply than implied, e.g., by the original Madau plot.
The discovery of larger samples of high-$z$ SNe~Ia by forthcoming
observational projects should yield strong constraints on the progenitor 
models and the SFH (see Gal-Yam \& Maoz 2003 for full details).   
In the previous section we have demonstrated that if SNe~Ia produce most 
of the iron in galaxy clusters, and the stars in clusters formed 
at $z\sim2$, the SN~Ia delay 
time must be {\it lower} than $2$~Gyr. If so, 
then the Lanzetta et al. (2002) SFH is inconsistent with the data
presented here (Figure 3).

\begin{figure}
\centering
\includegraphics[height=9cm]{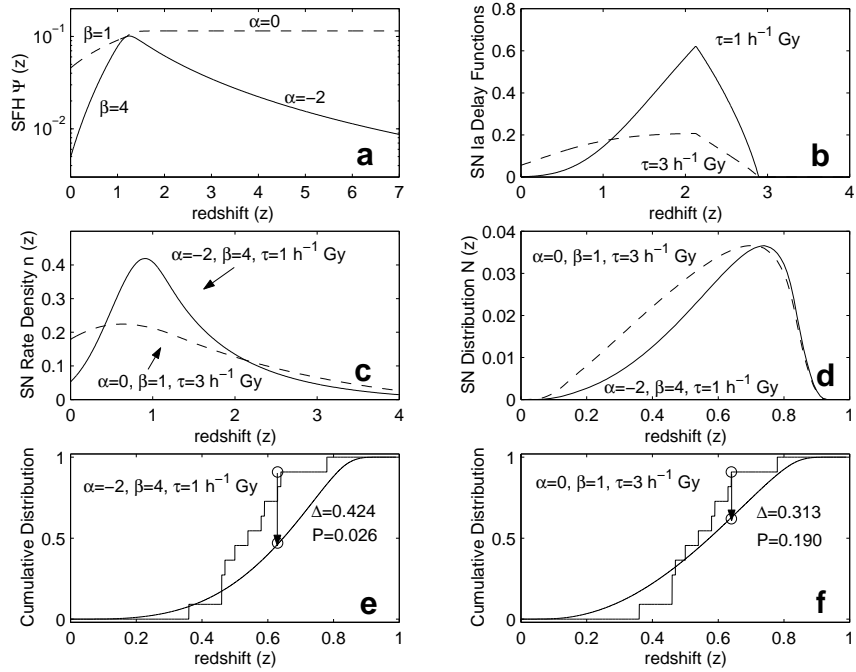}
\caption{Illustration of the modeling and comparison to data. 
The SFH ($\Psi(z)$) is modelled by two power laws,
smoothly joined at $z=1.2$. We denote the high-$z$ index with $\alpha$ and
the low-$z$ index with $\beta$. Panel a shows two examples, 
a ``Madau'' SFH, with a peak at $z=1.2$ (solid curve), 
and a shallower model (dashed curve) reflecting the proposed modifications 
by Cowie et al. (1999) and Steidel et al. (1999). 
Panel b shows two examples of 
the expected SN~Ia rate density following a brief
burst of star formation. These delay functions are 
calculated using the prescription of
Madau, Della Valle, \& Panagia (1998), with characteristic exponential 
delay times of $\tau=1 h^{-1}$ Gyr (solid) and $\tau=3 h^{-1}$ Gyr (dashed). 
For display purposes, an arbitrary redshift of $z=3$ has been chosen for
the burst of star formation.
SFH models are convolved with a delay function, and the
resulting SN rate densities $n(z)$ for a ``Madau'' SFH with $\tau=1 h^{-1}$ Gyr
(solid) and a ``Cowie-Steidel'' SFH with $\tau=3 h^{-1}$ Gyr (dashed) 
are shown in panel c. Panel d shows the predicted 
SN distributions, $N(z)$, for the models of panel c,
in a survey with the same observational parameters of the SCP search. 
KS tests show that the cumulative version of $N(z)$ from
a model combining a ``Madau'' SFH with a typical delay time of $\tau=1 h^{-1}$
Gyr (panel e) is ruled out by the data,  while a model with
``Cowie-Steidel'' SFH and $\tau=3 h^{-1}$ Gyr is consistent with the data
(panel f). This ``forward modelling'' type of analysis is more powerful than
the derivation of an observed SN rate at a mean $z$ from the data (e.g., 
Pain et al. 2002; Tonry et al. 2003) and comparison to predictions 
(as in panel c), since no binning over redshift is carried out.
Vertical axis units are arbitrary in panels a-d.}
\label{fig:2}       
\end{figure}

\begin{figure}
\centering
\includegraphics[height=9cm]{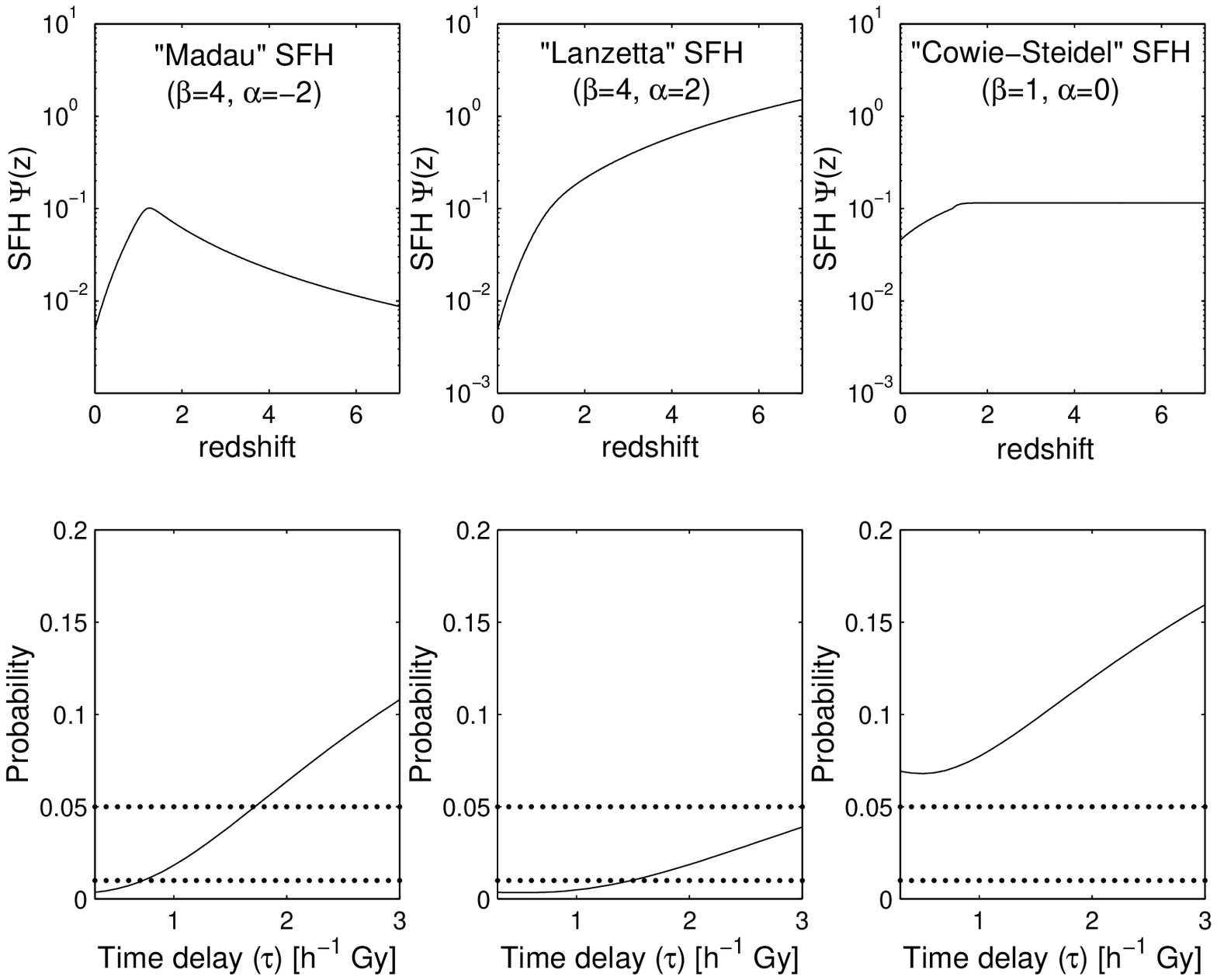}
\caption{Probability of SN~Ia time delay values $\tau$, given the data,
for particular SFH models. 
Assuming the SFH models shown in the upper panels, we can constrain
the allowed values of $\tau$ by the probability derived from the
KS test (lower panels). Points below the upper and lower dotted lines 
are ruled out at $95\%$ and $99\%$ confidence, respectively.}
\label{fig:3}       
\end{figure} 

\section{Conclusions}

The large mass of iron in clusters, combined with the low SN~Ia rate
we have measured in $z\sim1$ clusters, require that, if SNe~Ia are the 
dominant iron source, their delay time must be short. This would cast doubt,
e.g., on the double degenerate models. The observed redshift distribution 
of field SNe~Ia implies a slowly changing field SNR(z) for SNe~Ia. For several
popular star-formation histories, this then indicates a long delay time. Taken
together, these constraints may suggest that the iron in clusters is from 
core-collapse SNe, from an early stellar population with a top-heavy IMF.

\printindex
\end{document}